\documentclass[conference]{IEEEtran}
\IEEEoverridecommandlockouts
\usepackage{cite}
\usepackage{amsmath,amssymb,amsfonts}
\usepackage{algorithmic}
\usepackage{graphicx}
\usepackage{textcomp}
\usepackage{xcolor}
\usepackage{booktabs}
\usepackage{url}
\def\BibTeX{{\rm B\kern-.05em{\sc i\kern-.025em b}\kern-.08em
    T\kern-.1667em\lower.7ex\hbox{E}\kern-.125emX}}
\begin{document}

\title{TPM-Based Continuous Remote Attestation and Integrity Verification for 5G VNFs on Kubernetes}
\author{
\IEEEauthorblockN{Al Nahian Bin Emran, Rajendra Upadhyay, Rajendra Paudyal, Lisa Donnan, Duminda Wijesekera}
\IEEEauthorblockA{
\text{Mason Innovation Labs, 
George Mason University, Arlington, VA, 22201, USA}\\
\text{\{abinemra $\mid$ rupadhya $\mid$ rpaudyal $\mid$ ldonnan $ \mid$ dwijesek@\}gmu.edu}
}
}


\maketitle

\begin{abstract}
In the rapidly evolving landscape of 5G technology, the adoption of cloud-based infrastructure for the deployment of 5G services has become increasingly common. Using a service-based architecture, critical 5G components, such as the Access and Mobility Management Function (AMF), Session Management Function (SMF), and User Plane Function (UPF), now run as containerized pods on Kubernetes clusters. Although this approach improves scalability, flexibility, and resilience, it also introduces new security challenges, particularly to ensure the integrity and trustworthiness of these components. Current 5G security specifications (for example, 3GPP TS 33.501~\cite{etsi_ts_133501_v1860}) focus on communication security and assume that network functions remain trustworthy after authentication, consequently lacking mechanisms to continuously validate the integrity of NVFs at runtime. To close this gap, and to align with Zero Trust principles of “never trust, always verify”, we present a TPM 2.0-based continuous remote attestation solution for core 5G components deployed on Kubernetes. Our approach uses the Linux Integrity Measurement Architecture (IMA) and a Trusted Platform Module (TPM) to provide hardware-based runtime validation. We integrate the open-source Keylime framework (which natively provides node-level attestation) with a custom IMA template that isolates pod-level measurements, allowing per-pod integrity verification \cite{schear2016bootstrapping}. A prototype on a k3s cluster \cite{k3s_docs} (a lightweight CNCF-certified Kubernetes distribution consisting of 1 master, 2 worker nodes) was implemented to attest to core functions, including AMF, SMF and UPF. The experimental results show that the system detects unauthorized modifications in real time, labels each pod’s trust state, and generates detailed audit logs. This work provides hardware-based continuous attestation for cloud native and edge deployments, strengthening the resilience of 5G as critical infrastructure in multi-vendor and mission-critical scenarios of 5G.
\end{abstract}


\section{Introduction}
\label{sec:introduction}

The rollout of 5G networks has embraced cloud-native principles, with core virtual network functions (VNFs) such as the Access and Mobility Management Function (AMF), Session Management Function (SMF), and User Plane Function (UPF) being deployed as containerized microservices on Kubernetes clusters. This cloud-native transition enables scalability and agility but also expands the attack surface of the 5G core~\cite{toheeb2025impact}. In particular, running critical control-plane and user-plane functions in general-purpose cloud environments raises concerns about their runtime integrity and protection against advanced persistent threats or insider attacks. The 3GPP’s primary 5G security specification (TS 33.501)~\cite{etsi_ts_133501_v1860} defines a comprehensive security architecture for 5G systems, including mutual authentication and interface protection, but does not specify any mechanism to continuously validate the integrity of the software running network functions or utilize hardware trust anchors~\cite{etsi_ts_133501_v1860}. According to these specifications, once a VNF is authenticated and its connections are secured (e.g., using TLS), the current standards assume the VNF remains trustworthy; an assumption that may not hold against runtime compromises or kernel-level malware.

This challenge becomes even more critical in sensitive domains such as defense, aviation and mission-critical communications. In these scenarios, 5G is increasingly being deployed in secure cloud environments, private data centers, or in localized edge infrastructures to support applications such as command-and-control, air traffic communications, and battlefield connectivity. Any compromise in the run-time integrity of core functions in such environments could lead not only to service disruption but also to severe safety and security risks. Therefore, continuous integrity verification and attestation are essential wherever 5G is used to support critical infrastructure or national security operations.

To address this gap, we propose a TPM-based continuous remote attestation and integrity verification framework for 5G network functions. Our work learned from the dissertation of Piras~\cite{piras2022tpm}, while extending it to support Kubernetes-based VNFs of the 5G core such as AMF, SMF, UPF, etc. Remote attestation is a security technique in which a remote verifier challenges a prover machine to furnish evidence (often signed measurements) of its software state, allowing detection of unauthorized changes. A hardware Trusted Platform Module (TPM) provides a root of trust, securely storing cryptographic keys (Key management) and platform state measurements. Using TPM shielded capabilities, one can obtain cryptographic evidence of the integrity of the platform. In our approach, each 5G core node is equipped with a TPM 2.0, and Linux \emph{Integrity Measurement Architecture (IMA)} is used to record hashes of executables and critical files as they are loaded into both the host and container contexts. We integrate Keylime, an open source CNCF (Cloud Native Computing Foundation \cite{cncf_website}) project for scalable trust management and continuous monitoring. Keylime uses TPM 2.0 and IMA to implement a remote attestation platform for runtime integrity monitoring. It automates the process of bootstrapping hardware-rooted trust and continuously verifying measured state, alerting a party concern with cybersecurity management if the system deviates from an expected integrity baseline. Using Keylime’s framework and extending it for container-granular measurements, this solution continuously monitors the integrity of both the host platform and every 5G core pod, providing a layered trust model. To summarize, our contributions are as follows: (i) we re-implemented the approach of Piras~\cite{piras2022tpm} because no source code was publicly released, adapting it for a Kubernetes-based 5G core deployment; (ii) we extend the Keylime framework beyond its default node level scope by introducing a pod-aware IMA template and the accompanying verifier logic that parse Kubernetes cgroup (control groups) paths to bind measurements to Pod UIDs like AMF, SMF, UPF, and other core functions; and (iii) we developed a working prototype on a k3s cluster that demonstrates real-time detection of runtime integrity in 5G network function pods.

 \section{Related Work}
\label{sec:relatedWork}
Remote attestation has long been studied in NFV (Network Functions Virtualisation) and 5G systems as a means to enforce trust at runtime. Benedictis et al. introduce a Trust Monitor for ETSI NFV that continuously assesses the integrity of NFVI hosts and VNFs from the MANO (Management and Orchestration) domain, with a centralized verifier and modular “attestation drivers” to support heterogeneous RA (Remote Attestation) technologies \cite{de2019proposal}. Their prototype targets Security-as-a-Service and emphasizes scalability and vendor-agnostic integration, while noting challenges such as whitelist maintenance and integrating verifier logic across workflows. They also discuss container-aware IMA usage to distinguish measurements on shared hosts through per-container attribution in the IMA log. Building on the same TPM/IMA foundations, our approach moves the trust boundary down to the Kubernetes pod for 5G NFs attest and putting attestation details directly into cluster orchestration so that violations can automatically trigger remediation (e.g., eviction and restart).
In \cite{oliver2021trust}, Oliver explores embedding remote attestation into 5G/6G systems, extending trust from hardware up through containers and VMs, and coordinating identities and TPM quotes across core and edge domains . The proposal includes a Remote Attestation Service (RAS) integrated with the MANO stack and introduces the concept of trust slicing, where only attested entities gain unrestricted slice access. It also discusses practical concerns such as multi-TPM identity binding and how MANO components can consume TPM quotes to control admission. While Oliver provides the high-level architectural vision and PoC(Proof of Concept) scenarios, our work operationalizes continuous runtime RA inside a Kubernetes-managed 5G Core by producing per-pod trust states and enforcing policy-driven remediation when integrity deviates.

Outside of 5G/NFV architectures, there has been notable progress on general frameworks for continuous remote attestation using TPMs and IMA. Jordi et al. describe Keylime, a widely used open-source system that provides measured boot and continuous integrity monitoring for cloud nodes with TPM 2.0 as the root of trust. Keylime consists of a tenant (policy agent), a verifier service, and an attestation agent on each prover node, automating the collection of TPM quotes and IMA logs, comparing them against whitelists, and triggering alerts or actions if anomalies are detected \cite{thijsman2024trusting}. Margie et al. conducted an empirical study of Keylime in cloud deployments, showing that it can detect real-world attacks including ransomware, rootkits, and file tampering while also identifying blind spots such as unmeasured paths and false positives caused by system updates \cite{ruffin2025towards}.
The utility of TPM-based attestation has also been demonstrated in telecom and timing-critical infrastructures. Berbecaru et al. present a framework that employs TPM-based RA to protect a Time Distribution Network (TDN) used for telecom clock synchronization. They equipped White Rabbit PTP devices with TPMs (or vTPMs) and used Keylime to periodically verify daemon software and configuration, detecting subtle integrity attacks that would otherwise go unnoticed. Their experiments show that trusted computing significantly improves the robustness of time sync services, illustrating that TPM, IMA and remote verifier is effective not only for cloud servers but also embedded telecom devices \cite{berbecaru2023mitigating}.

Beyond these targeted deployments, the literature also surveys the wider landscape of attestation. Several works demonstrate remote attestation with different hardware and software approaches \cite{seshadri2004swatt, kong2014pufatt,tan2015remote}. Some leverage additional hardware features, such as Intel TXT \cite{intel2014txtguide}, while others especially in embedded systems explore software-only attestation without relying on dedicated hardware \cite{seshadri2004swatt}. Comprehensive surveys cover attestation schemes across cloud, IoT, and critical infrastructures \cite{sfyrakis2020survey}. More recently, in IoT environments, research has shifted toward collective remote attestation (CRA), which enables scalable verification of large networks of devices \cite{ambrosin2020collective}.
This work provides continuous, pod-level attestation of 5G core network functions with policy-driven remediation integrated into Kubernetes, thereby aligning runtime trust enforcement with the operational needs of cloud-native 5G.
 
\section{Motivation and Threat Model}
\label{sec:motivation&threatModel}

The 5G core network forms the backbone of emerging telecommunication networks, handling subscriber authentication, mobility management, session establishment, and traffic routing. Compromising these core functions can have devastating consequences, from large-scale denial of service to interception of sensitive data. As operators deploy 5G core VNFs on cloud infrastructure (often using container orchestration like Kubernetes), new security challenges emerge. Cloud deployments are susceptible to threats such as container escapes and host privilege escalation \cite{zhou2023container,Avrahami_2024}, supply chain attacks through compromised images \cite{mounesan2023exploring}, and insider threats that exploit misconfigurations \cite{gunasekhar2015mitigation}. Traditional security controls like network segmentation and static image scanning, while necessary, may not be sufficient to detect if a running VNF has been subverted (e.g., via injection of malicious code at runtime). Moreover, current 5G security standards do not yet mandate run-time attestation or integrity verification of network functions. (3GPP TS 33.501 release 18~\cite{etsi_ts_133501_v1860}) ensures that the network functions mutually authenticate and communicate through secure channels, but implicitly trust operating environment of the VNF once the VNF is initially authenticated. In practice, this means that if an attacker manages to compromise the software of an VNF (for example, by exploiting a zero-day vulnerability in the VNF container or underlying OS), the 5G security architecture has no built-in method to detect this compromise as long as the attacker does not breach the communication security. This lack of hardware-based trust and runtime integrity validation in the standard leaves a blind spot in the defense of 5G core networks.

Industry and academia have begun to recognize this gap. Best-practice guidelines for cloud native telecommunication deployments advocate for hardware-based trust roots (like TPMs or hardware security modules) and continuous monitoring of system integrity~\cite{de2019proposal,mavenir2021openran }. The motivation for our work is to bring these principles into the context of the 5G core. The objective of the paper is to provide continuous assurance that each core network function pod is running untampered code, by dynamically measuring and verifying its runtime state against a known-good reference. This assurance is especially important in scenarios like multi-vendor 5G cores or sensitive deployments (e.g. military, aviation, or other mission-critical 5G setups) where some VNFs might be supplied by third parties or operated in untrusted environments. In these scenarios, relying solely on compliance with standards or network layer security is insufficient, and a compromised VNF could act maliciously while still appearing \emph{normal} on the network. Our approach addresses the following threat model: an adversary may attempt to modify or inject executable code or libraries into a running 5G core VNF pod (through container breakout, exploiting a vulnerable service in the pod, or using a malicious insider with access to the node). We assume the underlying host’s secure boot and the TPM 2.0 establish a root of trust at boot time, and that the attacker has not physically tampered with the TPM. We assume that the secure boot of the underlying host and TPM 2.0 establish a root of trust at boot time and that the attacker has not physically tampered with the TPM. We also assume that the Kubernetes platform itself and the host OS kernel are up-to-date with IMA support enabled. Within this model, the goal is to continuously detect any unauthorized change in the software loaded and running in the core VNFs. Ideally, detection should be fast enough to enable automated mitigation (such as isolating or restarting a compromised pod) before significant damage is done.

In summary, our motivation is to fill the security gap by enforcing continuous runtime trust verification for core 5G network functions. This raises the security posture of the 5G system from one-time verification (at startup or deployment) to ongoing verification, aligned with zero-trust principles. Using a TPM-based attestation, we base this verification in hardware, making it difficult for an attacker to spoof the integrity evidence. Furthermore, by extending attestation to cover both the host platform and individual 5G core pods, our approach provides a layered trust model that strengthens the overall resilience of 5G deployments. The next section reviews related work and existing solutions that inform our design.

\section{Background}
\label{sec:background}

This section summarizes the core Trusted Computing technologies that underpin our 5G core attestation framework: the Trusted Platform Module (TPM), the Linux Integrity Measurement Architecture (IMA), and the remote attestation workflow as implemented by the Keylime framework.

\subsection{Trusted Platform Module 2.0}
\label{ssec:tpm2.0}

The \emph{Trusted Platform Module (TPM)} is a dedicated hardware security module that provides cryptographic primitives and a hardware root of trust. TPM 2.0 exposes a set of \emph{Platform Configuration Registers (PCRs)} that hold cumulative digests of the measured system components. During boot, each stage (firmware, BIOS, boot loader, kernel) is measured and extended into PCRs, establishing a verifiable chain of trust. In Linux systems, this chain is extended to runtime using the IMA subsystem, which uses PCR\-10 to hold runtime integrity measurements.

The TPM contains an Endorsement Key (EK), provisioned by the manufacturer, and one or more Attestation Keys (AKs). The AK is a non-migratable signing key used for quoting PCR values during attestation. To preserve privacy, the EK itself is not used directly for signatures; instead, the AK is certified by the EK or an external Privacy CA. This ensures that TPM quotes can be cryptographically validated as originating from a genuine TPM without exposing the EK itself.

\subsection{Linux Integrity Measurement Architecture}
\label{sec:LinuxIMA}

The \emph{Linux Integrity Measurement Architecture (IMA)} is a kernel subsystem designed to monitor and protect system integrity by recording hashes of files and executables as they are accessed. IMA maintains a Measurement List (ML), an append-only log that records each measured event. Each entry in the ML consists of a template defined by the system policy, which typically includes fields such as file hashes, pathnames, and signatures. Popular templates include \texttt{ima}, \texttt{ima-ng}, and \texttt{ima-sig}. For each entry, the digest of the template fields is extended into PCR~10, thereby binding the ML to the TPM state.

IMA policies define what gets measured. For example, the policy may specify that all executables, kernel modules, or configuration files must be hashed before use. The ML is exported using \texttt{securityfs} (e.g., {/sys/kernel/security/ima/ascii\_runtime\_measurements}) and can be retrieved for attestation purposes. A remote verifier can recompute the hashes of ML entries and compare them against the PCR~10 value quoted by the TPM. If the recomputed value matches the TPM quote, the verifier is assured of the ML’s integrity. Then they can compare every entry against a whitelist of known-good digests to detect tampering or unauthorized software execution.

\subsection{Remote Attestation Workflow}
\label{sec:remoteAttestationWorkflow}

Remote attestation (RA) provides a means for a trusted verifier to evaluate the software state of a remote system (the attester). The process begins with the verifier sending a nonce-based challenge. The attester, using its TPM, produces a quote over selected PCR values (e.g., PCR~10) signed with its Attestation Key (AK/AIK). The attester also returns the current IMA Measurement Log (ML), which records hashes of files and executables accessed by the system. The verifier then performs three checks: (i) that the TPM quote signature is valid under the AK, (ii) that the ML entries re-hash correctly to the quoted PCR~10 value, and (iii) that the ML entries match a whitelist of approved values. If all checks succeed, the attester is \emph{deemed Trusted}; if any discrepancy is found (e.g., missing or unexpected entries, incorrect digests), the attester is \emph{considered untrusted}.

\begin{figure}[htbp]
  \centering
  \includegraphics[width=0.48\textwidth, height=0.33\textheight]{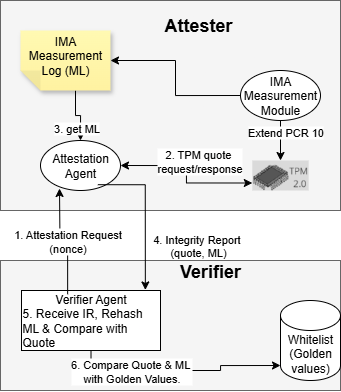} 
  \caption{TPM-backed Remote Attestation Workflow}
  \label{fig:tpm8}
\end{figure}

This workflow shown in Figure~\ref{fig:tpm8}) illustrates the six-step RA process. (1) The Verifier sends an attestation request including a nonce and PCR mask. (2) The Attestation Agent requests a quote from the TPM and receives an AIK-signed response. (3) The Agent retrieves the current IMA ML. (4) An Integrity Report containing the quote, ML, and nonce is returned to the Verifier. (5) The Verifier recomputes the digest of the ML to confirm that it matches the quoted PCR~10. (6) The Verifier then compares the validated quote and ML entries against golden values in its whitelist, producing a pass/fail trust state. This generic RA process forms the foundation that our Kubernetes-integrated framework extends to support per-pod integrity verification.

\subsection{Keylime Framework}
\label{ssec:kelime}

Keylime is an open source CNCF project that automates remote attestation based on TPM and IMA in distributed environments. Its architecture consists of four main components:
\begin{itemize}
  \item \textbf{Registrar}: Stores EK public keys, AK certificates, and node identities.
  \item \textbf{Verifier}: Periodically challenges attested nodes, validates TPM quotes and IMA MLs, and assigns trust states.
  \item \textbf{Agent}: Runs on each attested node, collects TPM quotes and MLs, and responds to verifier requests.
  \item \textbf{Tenant}: Configures attestation policies, whitelists, and initiates node registration.
\end{itemize}

\begin{figure}[htbp]
  \centering
  \includegraphics[width=0.48\textwidth, height=0.22\textheight]{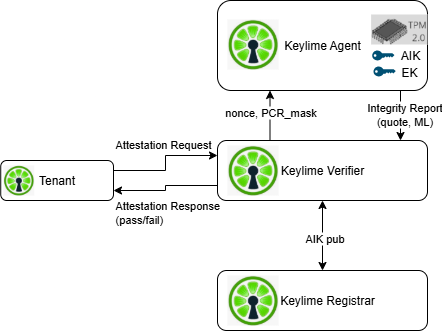} 
  \caption{Keylime continuous attestation workflow}
  \label{fig:tpm7}
\end{figure}

Keylime supports continuous attestation by polling agents at configurable intervals (default $\sim$2 seconds). It also implements a three-party bootstrap key-derivation protocol, enabling encrypted payload delivery to attested nodes. Keylime evaluates host integrity (TPM quote + IMA ML) at the node granularity, but it does not distinguish which container/pod generated a given IMA measurement. In our work, we extended Keylime with a custom IMA template to support pod-level attestation in Kubernetes. Figure~\ref{fig:tpm7} illustrates the workflow of Keylime enabling remote continuous attestation backed by a TPM. The \emph{Tenant} issues attestation requests according to policy, while the \emph{Verifier} challenges each \emph{Agent} with a nonce and selected PCR mask. The Agent, using its TPM, generates an AIK-signed quote on the PCR values and returns it with the current IMA measurement list (ML) as an \emph{Integrity Report}. The Verifier validates AIK against \emph{Registrar}, checks the consistency between ML and PCR values, and compares measurements against configured \emph{allow-lists}. A pass/fail decision is then returned to the Tenant, which can trigger higher-level orchestration actions. This cycle repeats periodically, enabling continuous verification of node integrity.

\section{System Model and Experimental Setup}
\label{sec:systemModel}

We implemented our attestation framework on a Kubernetes-based 5G core testbed. This section describes the cluster setup, the IMA configuration, the Keylime integration, and experimental validation.

\subsection{Cluster Setup}
\label{sec:clusterSetup}

\begin{figure*}[h]
  \centering
  \includegraphics[width=0.8\textwidth,height=0.37\textheight]{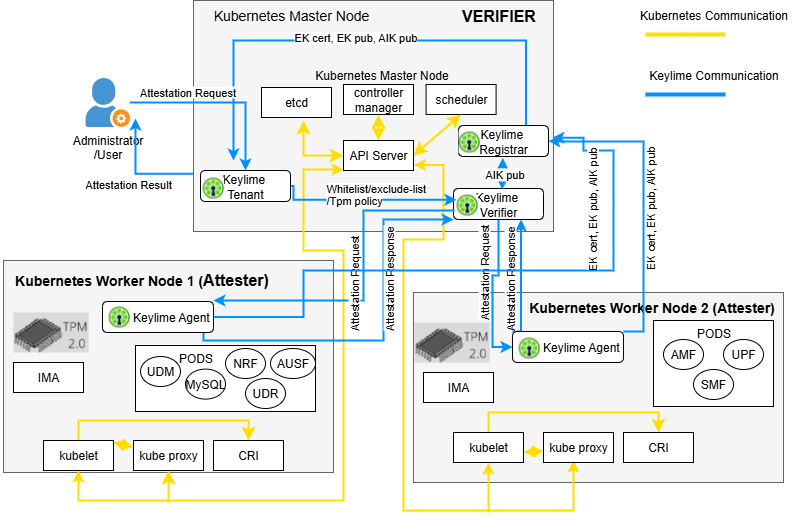}
  \caption{Basic Architecture of Kubernetes Cluster with Keylime Integration for Remote Attestation Utilizing TPM 2.0}
  \label{fig:tpm1-architecture}
\end{figure*}

\begin{figure}[htbp]
  \centering
  \includegraphics[width=0.48\textwidth]{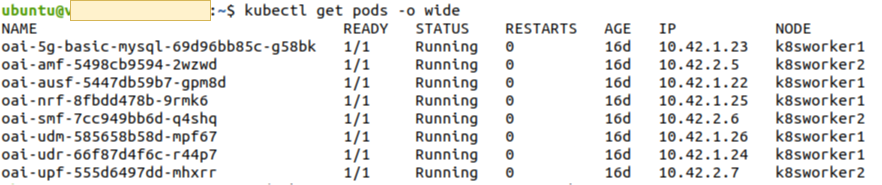} 
  \caption{Kubernetes Pod Deployment with Node and IP Mapping}
  \label{fig:tpm4}
\end{figure}

Figure~\ref{fig:tpm1-architecture} shows our end-to-end architecture of TPM/IMA-based attestation of a Kubernetes-hosted 5G core. The cluster was deployed using k3s~\cite{k3s_docs}(a lightweight CNCF-certified Kubernetes distribution) with one master node and two worker nodes. The 5G core functions were deployed from the OAI 5G Core implementation, containerized as Kubernetes pods~\cite{vilakazi2021evaluating}. All experiments were carried out on servers running Ubuntu 20.04.6 LTS with an IMA enabled custom Linux kernel 5.13.19. Every node in our cluster was provisioned with an Intel Xeon E-2278G CPU @ 3.40\,GHz (16 cores) and 8\,GB RAM, as well as a discrete TPM 2.0 device. The Kubernetes control-plane node (master node) runs the Keylime \emph{Verifier}, \emph{Tenant}, and the \emph{Registrar} alongside standard services (API server, scheduler, controller manager, and etcd). Two \emph{worker nodes} act as attesters. Worker~1 hosts support/control functions (\emph{MySQL}, \emph{NRF}, \emph{AUSF}, \emph{UDR}) while worker~2 hosts the 5G core functions \emph{AMF}, \emph{SMF}, and \emph{UPF}. Each worker includes a \emph{TPM~2.0} device, the Linux \emph{Integrity Measurement Architecture (IMA)}, and a \emph{Keylime Agent}. Kubernetes components such as \emph{Kubelet}, \emph{Kube-proxy}, and \emph{Container Runtime Interface (CRI)} manage pod execution. The Verifier continuously attests both node- and pod-level integrity. For completeness, we also include a topology view with pod\,$\leftrightarrow$\,node/IP mapping in Figure~\ref{fig:tpm4}. Figure~\ref{fig:tpm4} shows the 5G core network function pods deployed (AMF, SMF, UPF, etc.) with their assigned IP addresses and hosting nodes (k8sworker1, k8sworker2). This mapping allows the verifier to correlate pod-specific integrity measurements with the underlying worker node, enabling node-level and pod-level attestation.


\subsection{IMA Configuration and Template}
\label{ssec:IMAconfig}

By default, the IMA produces a single ML per node, which does not differentiate between host and container measurements. To enable pod-level granularity, we adopted the custom IMA template proposed by Piras~\cite{piras2022tpm}, extended with a \texttt{cgpath} field. This template records the cgroup path associated with each process, allowing the verifier to map ML entries to Kubernetes pod UIDs. In k3s, pod \texttt{cgroups} are consistently prefixed with \texttt{kubepods}, and the orchestrator path includes \texttt{/rancher/k3s}. Using this information, the verifier can extract pod identifiers from the ML and validate each pod independently.

\subsection{Keylime Integration}
\label{ssec:keylimeIntegration}

\begin{figure}[htbp]
  \centering
  \includegraphics[width=0.45\textwidth,height=0.27\textheight]{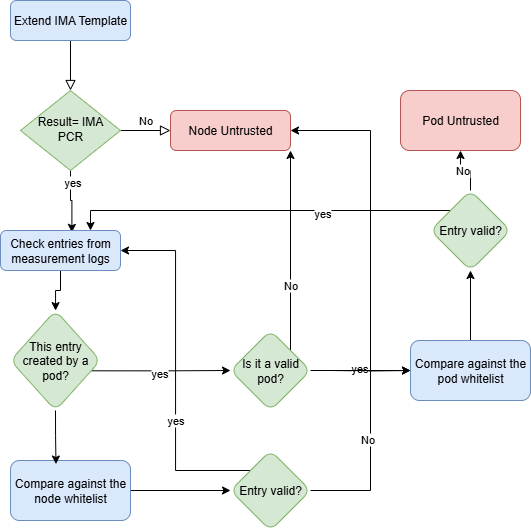} 
  \caption{TPM-backed remote attestation flow for Pod and Node Integrity Verification Using IMA Measurement Logs}
  \label{fig:tpm2}
\end{figure}

\begin{figure}[htbp]
  \centering
  \includegraphics[width=0.48\textwidth,height=0.15\textheight]{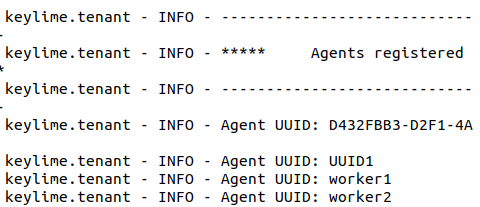} 
  \caption{Keylime Agent Registration in the Verifier/Registrar}
  \label{fig:tpm5}
\end{figure}

Each Agent registers its identity and TPM credentials with the Registrar (EK\textsubscript{cert}, EK\textsubscript{pub}, AIK\textsubscript{pub}). The Tenant supplies the Verifier with (i) a node allow list, (ii) pod-specific allow lists, (iii) optional exclude rules, and (iv) TPM policy (e.g., PCR selection). During each attestation cycle, the Verifier issues a nonce-based challenge; the Agent (1) asks the TPM to produce an AIK-signed quote over PCR~10, (2) collects the current IMA ML, and (3) returns an Integrity Report \{\emph{quote, ML, nonce}\} over mTLS. The Verifier validates the quote, proves ML\,$\leftrightarrow$\,PCR-10 consistency, and compares entries against the appropriate allow list (node or pod). Trust states are assigned as \emph{Start}, \emph{Trusted}, or \emph{Untrusted}. Figure~\ref{fig:tpm2} summarizes the attestation flow that illustrates the process of validating the run-time integrity in Kubernetes. Measurement entries are extended into IMA PCRs and then verified against node and pod whitelists. Entries originating from pods are mapped using pod identifiers and validated against pod-specific whitelists, while system-level entries are validated against the node whitelist. If discrepancies are found, the respective pod or node is marked as \emph{untrusted}. Figure~\ref{fig:tpm5} shows the Keylime tenant output showing the worker nodes (Agents) registered within the registrar. Each worker node is assigned a Universally Unique Identifier (UUID)(e.g., worker1, worker2), which is later used by the Verifier to track the attestation status. This registration step establishes cryptographic identities and prepares the system for continuous integrity verification of both nodes and pods.

\subsection{Pod Registration and Whitelists}
\label{ssec:PodRegistration+whiteLists}

\begin{figure}[htbp]
  \centering
  \includegraphics[width=0.48\textwidth]{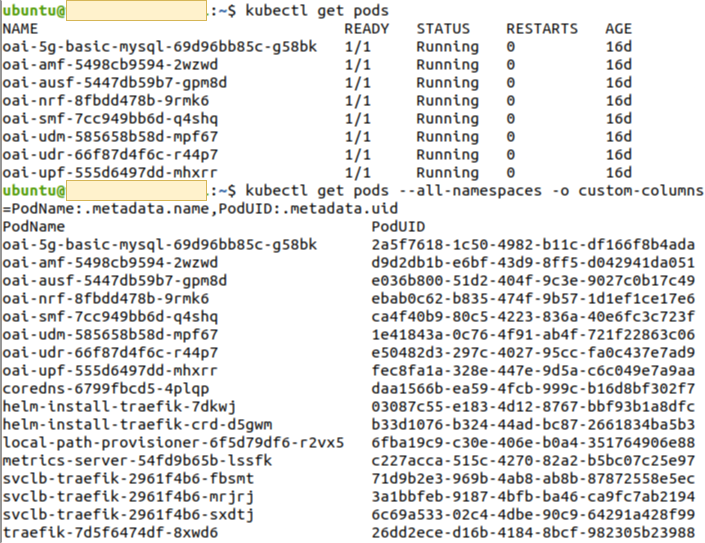} 
  \caption{Kubernetes Pod Deployment with Metadata and Pod UIDs}
  \label{fig:tpm3}
\end{figure}

Pod-level attestation requires registering the set of expected pods and their associated allow lists. The tenant provides this to the Verifier at the registration time. If an ML entry contains an UID for the pod that is not in the registered set, the node is marked \emph{untrusted} because an unknown pod is running. If the measurements of a registered pod deviate from its allow list, that pod is marked \emph{untrusted}, while the node may remain \emph{trusted}. This layered trust model prevents a single compromised pod from invalidating the entire node. Figure \ref{fig:tpm3} showing the core functions of 5G deployed alongside their unique Pod UIDs. These identifiers are extracted from the container control group path in the IMA logs and used by the verifier to link integrity measurements to specific pods for attestation.

\subsection{Selective Validation Experiments}

\begin{figure}[htbp]
  \centering
  \includegraphics[width=0.48\textwidth]{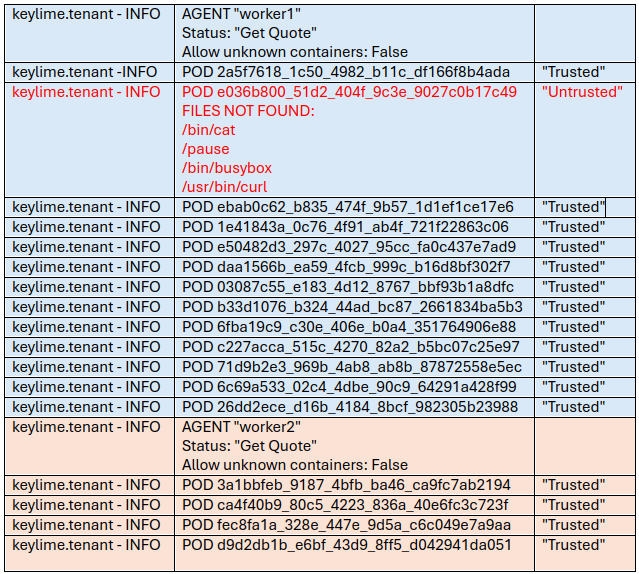} 
  \caption{Keylime Continuous Attestation Results with Pod Trust States and Whitelist Violations}
  
  \label{fig:tpm6}
\end{figure}

Figure~\ref{fig:tpm6} shows the Keylime tenant output for both Worker~1 and Worker~2 during a continuous attestation cycle. In this experiment, the Verifier was restricted to monitoring files only in /usr/bin (regex\verb!^(?!/usr/bin/).\*\$). Attesting an entire host or container image would otherwise produce a very large measurement log, so this controlled set-up was chosen to clearly demonstrate how unexpected files trigger runtime integrity violations while the overall node remains trusted. 

On Worker~1, most pods including \emph{MySQL}, \emph{NRF}, and \emph{UDR} are classified as \emph{Trusted}. However, the \emph{AUSF} pod is flagged as \emph{Untrusted}. The Verifier output lists discrepancies such as unexpected or missing binaries: \texttt{/bin/cat}, \texttt{/pause}, \texttt{/bin/busybox}, and \texttt{/usr/bin/curl}. These binaries were not part of the registered whitelist, so their appearance in the measurement log immediately caused the pod to fail attestation. In practice, \texttt{/pause} and \texttt{busybox} are helper containers automatically spawned by Kubernetes, while \texttt{cat} and \texttt{curl} were executed manually inside the pod during testing. Because none of these binaries were whitelisted, the attestation policy flagged the pod as untrusted. This experiment shows how both routine Kubernetes helpers and ad-hoc command executions are caught when they deviate from the expected whitelist.

On Worker~2, all deployed pods including AMF, SMF, and UPF remain \emph{Trusted}. This indicates that when pods conform to their registered whitelists, the system maintains both node-level and pod-level trust. By distinguishing between compliant and noncompliant pods, the framework enables fine-grained remediation strategies (e.g., evicting only the compromised pod \emph{AUSF} while keeping the rest of the system operational.

\subsection{Results Summary}
\label{ssec:summary}

The experimental results shown in Figure~\ref{fig:tpm6} show that the extended Keylime framework is capable of distinguishing between trusted and untrusted pods while preserving the integrity of the node level. In the observed cycle, Worker~1 hosted multiple pods, most of which (\emph{MySQL}, \emph{NRF}, \emph{UDR}) remained in a state of \emph{Trusted}, while the pod of \emph{AUSF} was marked as \emph{Untrusted}. The Verifier output explicitly listed binaries such as \texttt{/bin/cat}, \texttt{/pause}, \texttt{/bin/busybox}, and \texttt{/usr/bin/curl} that were not present in the pod’s allow list, offering clear evidence of the violation. Worker~2, running \emph{AMF}, \emph{SMF}, and \emph{UPF}, maintained a fully trusted state, showing that compliant pods can operate unaffected even when another pod fails attestation. These results highlight two key facts. First, the attestation log provides administrators with forensic visibility into the exact cause of the violation, enabling them to trace which binaries or processes led to integrity failures. Second, the framework allows administrators to enforce remediation policies: depending on configuration, an untrusted pod may be shut down and rescheduled as a fresh pod, while uncompromised pods and nodes remain operational. This policy-driven response ensures that local failures do not escalate in to service-wide outages.  

In the context of telecommunication deployments, where 5G cores often involve multiple network functions running across distributed Kubernetes clusters, this layered trust model is particularly valuable. Runtime violations can be isolated to specific pods without undermining the trust in the entire node or cluster. This not only improves resilience and up-time, but also aligns with zero-trust principles by providing continuous validation of both host and pod integrity. By combining hardware-based attestation with pod-level granularity, the framework offers operators a practical mechanism to protect critical 5G core services against runtime tampering, while supporting policy-driven automated remediation and recovery that minimizes disruption to live network operations.

\subsection{Performance Overhead}
\label{ssec:performanceOverhead}

In addition to functional validation, we measured the performance overhead introduced by continuous TPM/IMA-based remote attestation in our Kubernetes-based 5G core deployment. Specifically, we evaluated CPU consumption on both worker nodes with and without the Keylime agent running. We report both the average and the 95th-percentile (p95) CPU utilization, where the average captures overall overhead and p95 highlights whether attestation introduces periodic spikes.

\begin{figure}[t]
  \centering
  \includegraphics[width=0.48\textwidth]{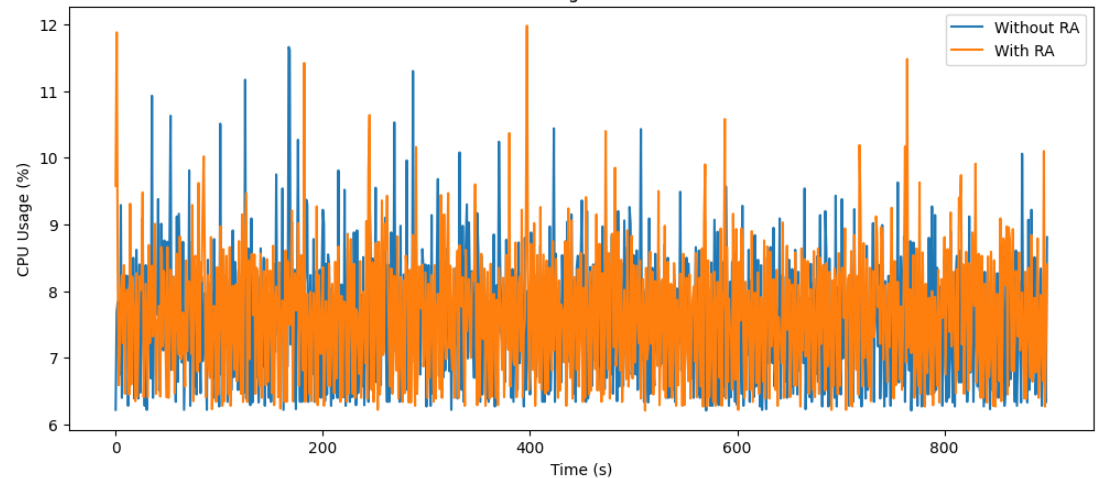}
  \caption{Worker Node~1 CPU utilization with and without RA.}
  \label{fig:worker1_cpu}
\end{figure}

\begin{figure}[t]
  \centering
  \includegraphics[width=0.48\textwidth]{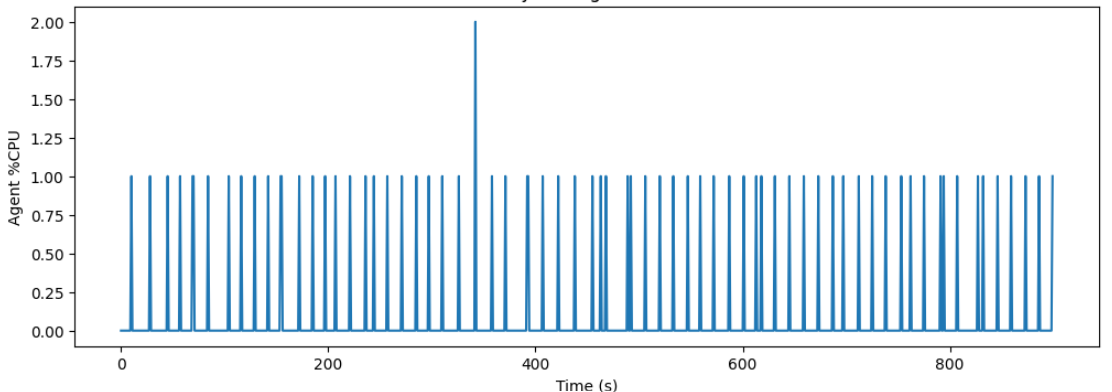}
  \caption{Keylime agent CPU usage on Worker Node~1.}
  \label{fig:worker1_agent}
\end{figure}

\begin{figure}[!t]
  \centering
  \includegraphics[width=0.48\textwidth]{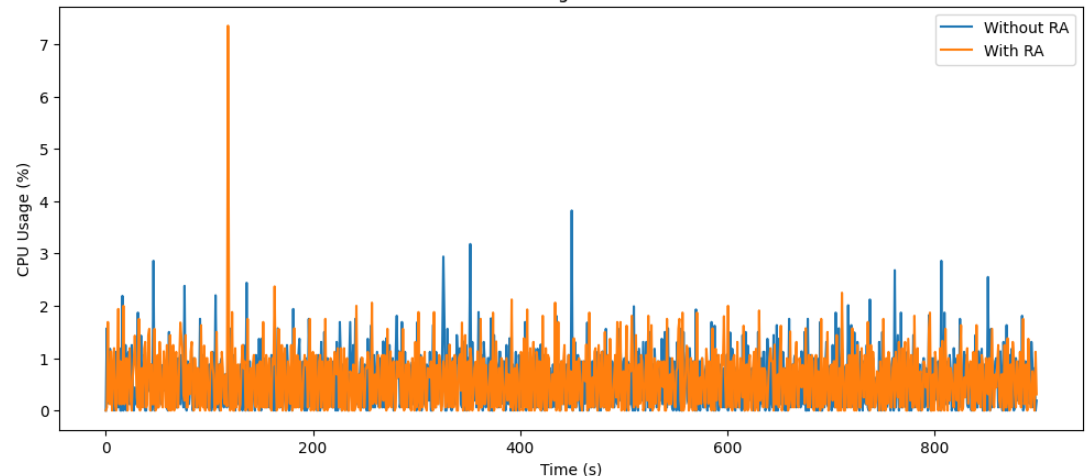}
  \caption{Worker Node~2 CPU utilization with and without RA.}
  \label{fig:worker2_cpu}
\end{figure}

\begin{figure}[!t]
  \centering
  \includegraphics[width=0.48\textwidth]{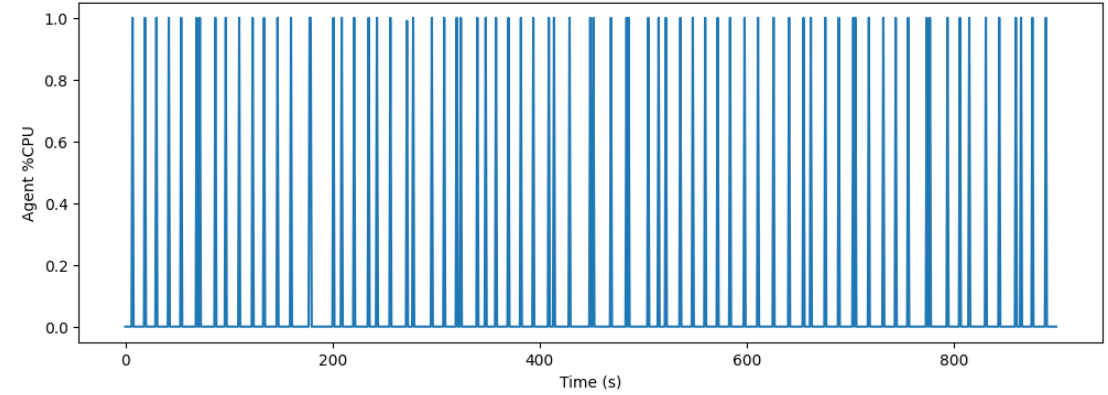}
  \caption{Keylime agent CPU usage on Worker Node~2.}
  \label{fig:worker2_agent}
\end{figure}

For Worker~1, the results show negligible performance impact. As illustrated in Fig.~\ref{fig:worker1_cpu}, the average node-level CPU utilization increased by only 0.04\% compared to baseline, while the p95 remained unchanged. Fig.~\ref{fig:worker1_agent} further shows the CPU usage of the Keylime agent itself, which consumed an average of 0.08\% CPU with periodic spikes of approximately 1\%-2\% at quote intervals, consistent with the lightweight nature of TPM quoting and IMA log transfer. Worker~2 results are consistent with those of Worker~1. As shown in Fig.~\ref{fig:worker2_cpu}, the average CPU consumption rose by only 0.0023\% compared to baseline, and the p95 increased by less than 0.01\%, well within measurement noise. Fig.~\ref{fig:worker2_agent} shows that the Keylime agent on Worker~2 averaged just 0.083\% CPU, with small periodic spikes of about 1\% during quote intervals.

Table ~\ref{tab:cpu_overhead} summarizes the quantitative results for both workers. Together, these results demonstrate that continuous pod-level attestation imposes negligible performance overhead on Kubernetes worker nodes. Since attestation runs in the background and does not sit on the data path, it does not add delay to 5G signaling or user traffic. The only time service is affected is when a violation is detected and, based on policy, the system may evict or restart a pod. This is a deliberate enforcement action rather than normal latency overhead. Even when we measured at both the node-wide level and the individual agent-process level across multiple workers, the additional CPU cost of attestation was so small that it was indistinguishable from normal background fluctuations, confirming the practicality of continuous attestation in production-grade 5G core deployments.

\begin{table}[!t]
\centering
\caption{CPU Utilization With and Without Remote Attestation (RA)}
\label{tab:cpu_overhead}
\begin{tabular}{lccc}
\toprule
\textbf{Metric} & \textbf{Baseline (no agent)} & \textbf{With RA} & \textbf{ Overhead} \\
\midrule
\multicolumn{4}{c}{\textbf{Worker 1}} \\
\midrule
Node CPU (avg \%)   & 7.569 & 7.613 & +0.044 \\
Node CPU (p95 \%)   & 9.161 & 9.150 & -0.011 \\
Agent CPU (avg \%)  & --    & 0.080 & --     \\
\midrule
\multicolumn{4}{c}{\textbf{Worker 2}} \\
\midrule
Node CPU (avg \%)   & 0.578 & 0.580 & +0.002 \\
Node CPU (p95 \%)   & 1.621 & 1.630 & +0.010 \\
Agent CPU (avg \%)  & --    & 0.083 & --     \\
\bottomrule
\end{tabular}
\end{table}

\subsection{Relationship to the Threat Model and Attack Coverage}
\label{ssec:relatingThreatModel+AttackCoverage}

The experimental results shown in Section~\ref{ssec:summary} directly validate the assumptions of our threat model (Section~\ref{sec:motivation&threatModel}). We assumed that an adversary may attempt to modify or inject binaries within a running 5G core NF pod by exploiting containers that escape, malicious insider activity, or supply chain manipulation. In practice, when such events occurred in our prototype, for example, the execution of unexpected binaries like /bin/cat or /usr/bin/curl inside the AUSF pod, Keylime immediately flagged the pod as Untrusted. The Kubernetes runtime can then remove and restart the compromised pod with a fresh Pod UID, effectively removing the persistence of the attacker and restoring a clean state. This shows that our attestation framework not only detects violations, but also supports the automated recovery of the policy base, aligning the implementation results with the threat model.

Container escape and privilege escalation attacks (e.g., runC CVE-2019-5736) have been extensively studied in the context of container runtime vulnerabilities~\cite{Avrahami_2024}. In CVE-2019-5736, an attacker accessing a container can cause runC to execute its own host binary by using procfs (e.g. \texttt{/proc/self/exe}) and then overwrite that host binary by running \texttt{/proc/<runc-pid>/exe}, thereby achieving subsequent execution of root-level code on the host when runC is invoked again (e.g., using docker exec). This can be triggered either by using a malicious image (entrypoint/shebang redirect to \texttt{/proc/self/exe}) or during docker execution into a container where an expected binary was replaced, or also by using a malicious shared library (for example, libseccomp) that runs at load time. The upstream fix makes runC run itself again from a sealed in-memory copy, so that the runC binary on the host disk cannot be overwritten~\cite{Avrahami_2024}. Using our system in this scenario, if an attacker tampers with pod-resident binaries or injects additional tools (e.g. curl, nc) as part of the exploit staging, our pod-level IMA policy detects the hash deviation, and Keylime marks the pod untrusted. If the attacker escalates to overwrite a host runtime binary (for example by invoking /usr/sbin/runc) to gain persistence, the node’s IMA baseline diverges. Then, our node-level attestation will flag the worker node as untrusted. In summary, pod-scoped changes are caught by per-pod whitelists and host-scope changes are caught by node attestation. Although modern runC includes memfd mitigation (the patch that uses memfd\_create() to protect runC against overwrite attacks), our results show that TPM/IMA-anchored attestation adds an additional detection layer in this scenario.

Insider threats are another major concern in multi-tenant cloud environments. Gunasekhar et al.~\cite{gunasekhar2015mitigation} classify insider risks into four categories: (i) pure insiders such as employees or system administrators with extensive privileges; (ii) insider associates such as contractors, guards, or business partners with limited physical or system access; (iii) insider affiliates such as friends or family members who may gain access indirectly (e.g., using shared credentials); and (iv) outside affiliates such as external actors who exploit weak organizational controls such as unsecured wireless networks \cite{ashik2025reaperpulse,rahman2023pacman}. All of these categories can abuse their position to steal, modify, or leak organizational data. For telecommunication operators running multitenant 5G cores, such insider misuse could target sensitive signaling or subscriber data within network functions. Their mitigation strategy is data-centric, splitting trust by storing encrypted data in one cloud and keys in another, coupled with certificates and auditing. Our contribution addresses the complementary runtime dimension: insiders who attempt to execute ad hoc utilities (such as \texttt{curl} and \texttt{netcat}) to exfiltrate causes pod files or binaries to deviate from the configured IMA allowlist. Keylime detects these deviations, flags the pod as untrusted, and allows policy-driven remediation (eviction / restart). Thus, while Gunasekhar et al. mitigate at-rest data exposure, our attestation constrains insider actions at runtime with an additional detection layer against insider abuse in cloud-native telecommunications environments.

At the user space level, adversaries can use dynamic linker hijacking using a command such as \texttt{LD\_PRELOAD} or \texttt{/etc/ld.so.preload} to force a malicious shared object to load before legitimate libraries and hook critical functions such as \texttt{execve} or \texttt{readdir} for stealth, credential theft or command masking~\cite{wiz_dynamic_linker}. MITRE ATT\&CK formalizes this technique as Hijack Execution Flow: Dynamic Linker Hijacking (T1574.006), which enables persistence, privilege escalation, and defense evasion by interposing attacker code ahead of system libraries~\cite{mitre_ld_preload}. Observable artifacts include expected edits to \texttt{/etc/ld.so.preload} or unexpected library paths in process environments. In our framework, when an attacker abuses LD\_PRELOAD, they leave behind detectable footprints: changed preload files, suspicious new libraries, or unusual runtime access patterns. These changes cause mismatches in IMA’s integrity logs, which our attestation framework labels as tampering.  Any deviation results in the pod being marked \emph{untrusted}. Together, these runtime manipulations, whether occurring in the user space or in the kernel space, are exposed through TPM-anchored attestation and enforced by Keylime’s policy-driven orchestration. 

In summary, our evaluation demonstrates that TPM/IMA-based continuous attestation reliably flags tampering events across container escape, insider misuse, and runtime code injection scenarios, and thereby supports policy-driven remediation at both pod and node levels, directly aligning system behavior with the defined threat model mitigation.

\section{Relation to Zero Trust and Maturity Levels}
\label{sec:ZTA}

\begin{figure}[htbp]
  \centering
  \includegraphics[width=0.45\textwidth,height=0.18\textheight]{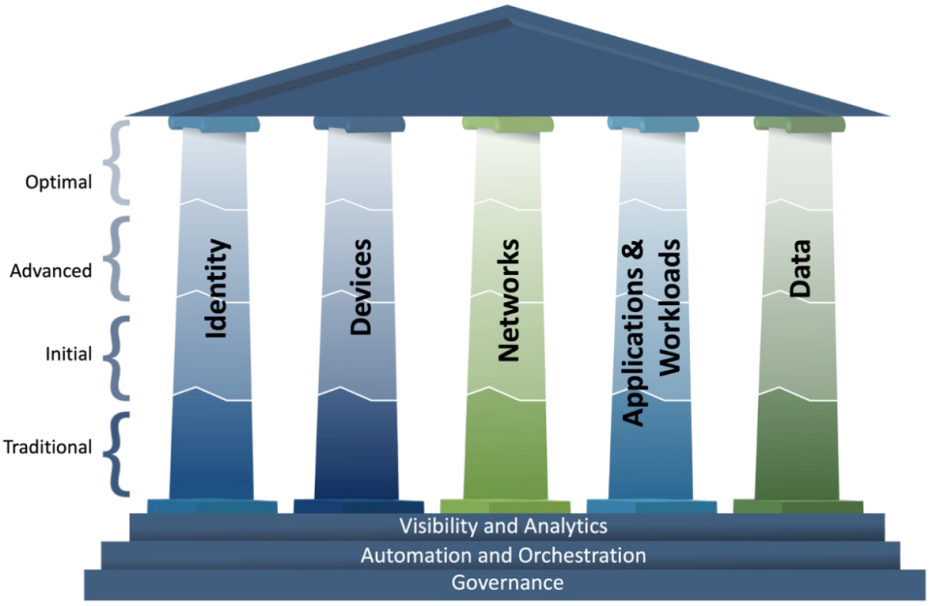} 
  \caption{Zero Trust Maturity Evolution~\cite{cisa2023ztmm}}
  \label{fig:zt_maturity}
\end{figure}

Zero Trust (ZT) is a security framework built on the principle of \emph{never trust, always verify}, requiring continuous validation of both identity and system posture before granting access to resources. The Cybersecurity and Infrastructure Security Agency (CISA) organizes Zero Trust into five security pillars: Identity, Devices, Networks, Applications \& Workloads, and Data, as illustrated in Fig.~\ref{fig:zt_maturity}~\cite{cisa2023ztmm}. Each pillar is assessed along the Zero Trust Maturity Model (ZTMM) (Fig.~\ref{fig:zt_maturity}), which progresses through four stages: \emph{Traditional} (manual, silo'ed and static), \emph{Initial}, \emph{Advanced} (increasing automation and cross-pillar visibility), and \emph{Optimal} (fully automated, adaptive, and enterprise-wide enforcement). The U.S. Department of War (DoW) Zero Trust Reference Architecture adopts a similar structure, underscoring its relevance for critical infrastructures such as 5G~\cite{freter2022department}.

Our approach supports some requirements of Zero-Trust principles. Rather than assuming that a network function remains trustworthy once authenticated, we enforce continuous verification of pod integrity and enable policy-driven remediation, such as eviction or restart of compromised instances. This capability ensures that trust is never static but continuously validated, consistent with Zero Trust architectures increasingly advocated for cloud-native and 5G systems. In terms of maturity, our framework goes beyond the \emph{Initial} stage to the \emph{Advanced} stages of CISA ZTMM~\cite{cisa2023ztmm}. It clearly exceeds the Initial stage by providing automated and continuous run-time attestation of workloads, rather than relying on static admission control or manual remediation. Through integration with Kubernetes, policy-driven actions such as pod eviction and restart are automatically triggered upon integrity violations, which aligns with the characteristics of the Advanced stage. In particular, our system advances the Applications \& Workloads and Devices pillars, because attestation is anchored in TPM hardware on worker nodes and extended to 5G core pods managed as workloads. The attack scenarios demonstrated in Section \ref{ssec:relatingThreatModel+AttackCoverage}, container escape (e.g., CVE-2019-5736), insider misuse of utilities, and LD\_PRELOAD-based runtime tampering can map directly onto zero trust assumptions that workloads and devices cannot be implicitly trusted. Our system shows how these threats can be addressed through TPM/IMA-based attestation and policy-driven remediation, thereby advancing Zero Trust adoption in a 5G core context. However, the framework does not yet reach the \emph{Optimal} stage, which CISA defines as requiring fully automated just-in-time lifecycles, dynamic least-privilege access, and cross-pillar interoperability with continuous monitoring. Thus, our contribution can be positioned within the Advanced maturity level for the relevant pillars, providing a practical step toward Zero Trust adoption in cloud native 5G cores. 
Although not shown here, our Kubernetes-mounted system can host other sensitive applications of Defense and Intelligence.
that require continuous monitoring and attestation against integrity and malicious code attacks. This is the primary reason for us to use the CISA and DoW ZTA nomenclature.


\section{Conclusion}
\label{sec:conclusion}

This paper presents a TPM-anchored continuous remote attestation framework for Kubernetes-based 5G core platforms. By extending Keylime with pod-aware IMA measurement policies, our prototype demonstrated that individual network functions can be monitored at runtime and that unauthorized modifications inside pods are promptly detected. The experimental results showed that when adversaries injected unexpected binaries or attempted insider-style tampering, the verifier flagged the affected pod as Untrusted, and Kubernetes remediation could evict and restart it with a fresh Pod UID. These outcomes directly validated our threat model, in which adversaries may compromise the run-time environment of 5G core functions while the TPM and secure boot remain trusted. Our work highlights the practical feasibility of integrating hardware-based trust into cloud-native telecommunications infrastructure. By requiring attestation at the pod level, operators can strengthen the security posture of 5G network functions against container escape, supply chain compromise, and insider abuse. Our approach also aligns with Zero Trust principles: rather than assuming that a network function remains trustworthy once authenticated, we continuously verify its integrity and enable policy-driven remediation, such as eviction or restart of compromised instances. This ensures that trust is never static, but continuously validated, consistent with Zero-Trust architectures increasingly advocated for cloud native and 5G systems~\cite{mavenir2021openran}.

Although our current work empirically validates the prototype, ongoing work includes formal verification of the attestation workflow. This could be achieved using symbolic protocol analysis tools (such as Tamarin Prover\cite{basin2025modeling}) to prove that a compromised VNF cannot forge trusted attestation evidence under our threat model. At the implementation level, deductive verification frameworks (e.g., Frama-C/WP \cite{blanchard2020introduction}) may be applied to critical parsing functions such as IMA log processing and pod UID extraction, guaranteeing memory safety and correctness. Pursuing these directions will provide complementary assurance, both empirical and mathematical, that the proposed framework can meet the trust requirements of next-generation 5G deployments. 
\bibliography{custom}
\bibliographystyle{IEEEtran}

\end{document}